\documentstyle[aps,prl,multicol,epsfig] {revtex}
\begin{document}

\newcommand{\lsmo}{La$_{0.7}$Sr$_{0.3}$MnO$_3$ }
\newcommand{\lsmaox} {La$_{0.7}$Sr$_{0.3}$Mn$_{1-\delta}$Al$_\delta$O$_3$ }
\newcommand{\lsmaothree}{La$_{0.7}$Sr$_{0.3}$Mn$_{0.97}$Al$_{0.03}$O$_3$ }
\newcommand{\lsmaosix}{La$_{0.7}$Sr$_{0.3}$Mn$_{0.94}$Al$_{0.06}$O$_3$ }

\title{Photoemission Line shape Study on
La$_{0.7}$Sr$_{0.3}$Mn$_{1-\delta}$Al$_\delta$O$_{3}$ ($\delta$ = 0, 0.03, 0.06) }
\author{Han-Jin Noh}
\address{School of Physics \& Center for Strongly Correlated Materials Research (CSCMR), Seoul National University, Seoul 151-742, South Korea}
\author{E.-J. Cho}
\address{Department of Physics, Chonnam National University, Kwangju, 500-757, South Korea}
\author{K.H. Kim\cite{khkim}, H.-D. Kim\cite{hdkim}, and S.-J. Oh\cite{sjoh}}
\address{School of Physics \& CSCMR, Seoul National University, Seoul 151-742, South Korea}
%\date{\today}
\date{  , 2001}
\maketitle

\begin{abstract}
We have studied the line shapes of ultraviolet photoemission
spectra of \lsmaox ($\delta$ = 0, 0.03, 0.06) systems to test the
extrinsic image charge screening effect on photoemission spectra
recently suggested by R. Joynt (Science {\bf 284}, 777 (1999)),
who argued that the photoemission spectrum near the Fermi energy,
specially for poorly conducting system, can be very different from
the intrinsic density of states because the outgoing electron has
probability of losing its kinetic energy due to the image force.
We tested this argument in real materials experimentally by
measuring the photoemission line shapes of \lsmaox systems, for
which all the requirements of this theory are satisfied and sample
resistivities change systematically. We found that experimental
photoemission spectra do not show the change of line shapes
expected from the extrinsic image force effect and we conclude
that the influence of this long range interaction is not so large
as suggested in that paper.

\end{abstract}


\begin{multicols}{2}
\section{Introduction}
\label{sec:1} Photoemission  Spectroscopy (PES) has been widely
considered as the most powerful tool for probing the electronic
structure of the occupied states in condensed matter systems.
Conventionally photoemission process is explained by a three-step
model\cite{hufner} which involves (1) the excitation of the
photoelectron, (2) its travel to sample surface, and (3) the
escape through the surface into the vacuum. This simple model,
though it is a purely phenomenological approach and has some
short-comings due to its semi-classical nature, has proved to be
quite successful in explaining many features in photoemission
spectra.

Recently, R. Joynt\cite{joynt} has suggested that more careful
consideration is needed in the third step, where the outgoing
electron experiences an attractive interaction due to its image
charge in the sample part during its travel from the sample to the
analyzer, so can lose its kinetic energy with some probability
distribution. Moreover it was argued that this long range
interaction effect should appear most drastically when the sample
resistivity is rather high (roughly $\rho_0 \gtrsim 0.1 m\Omega
cm$), which is based on his calculation results of energy loss
probability distribution and some fitting parameters taken from
experiments. The argument of Ref. 2 is so timely that many
photoemission spectroscopists pay attention to this assertion
because the currently much-studied materials such as high-T$_c$
superconductors\cite{supercond} and colossal magnetoresistance
(CMR) materials\cite{dessau} show so-called pseudogap features and
this theory looks like a master key to interpret these phenomena,
especially for the case of CMR manganites.

Though this argument  certainly contains some elements that should
be considered seriously, there are different opinions and opposing
views. D. S. Dessau {\em et al.}\cite{dessau2} gave a technical
comment that their angle-resolved photoemission (ARPES) of single
crystalline double layered manganite
La$_{1.2}$Sr$_{1.8}$Mn$_{2}$O$_7$ shows the momentum-dependent
pseudogap feature, which could not originate from the extrinsic
loss in Ref. 2 because the extrinsic loss should not change much
with momentum. Recently Y.-D. Chuang {\em et al.}\cite{chuang}
gave a new explanation about the pseudogap of this double layered
manganite that it originates from a short range charge/orbital
density wave enhanced by Fermi surface nesting. D. L.
Mills\cite{mills} followed up the approach of Ref. 2 with his own
calculations to give different results using the sum rule which
determines the total amount of energy loss probability, and
suggested that the extrinsic loss just shifts the kinetic energy
of all photoelectrons downward by roughly the same amount in the
usual photoemission experiments. Schulte {\it et
al.}\cite{schulte} presented several arguments, both theoretical
and experimental, that the zero energy loss probability $P_0$ is
not so small as was assumed in Ref. 2. They also obtained
$P(\omega)$ from their electron energy loss spectroscopy (EELS)
spectra of double layered manganite
La$_{1.2}$Sr$_{1.8}$Mn$_{2}$O$_{7}$ to calculate PES spectra and
compare them with the photoemission spectra. From these
comparisons they asserted this extrinsic effect cannot make
pseudogap-like feature in photoemission spectrum as long as $P_0$
is different from zero, and in most cases can be either neglected
or treated as a weak structureless background. In the recent
preprint\cite{joynt2}, R. Haslinger {\it et al.} gave quantitative
criterion when ohmic losses are important and pointed out the
layered manganite La$_{1.2}$Sr$_{1.8}$Mn$_{2}$O$_{7}$ is not a
good candidate for testing the theory of inelastic processes in
PES because its resistivity is too high and its crystal structure
is not cubic. The present responses to Ref. 2, both theory and
experiment, are somewhat critical. However, more experimental
results are needed. Since many strongly correlated materials lie
close to a metal-insulator phase transition and possess a large
resistivity, determining whether the argument of Ref. 2 is correct
becomes important. Furthermore, as far as we know, though a few
indirect evidences were reported as briefly described above, there
are no reports that give the direct experimental comparison and
analysis about the relation between the resistivity and line shape
of PES. In addition, previously studied system might be inadequate
for this test as claimed in Ref. 9. In this paper, we choose the
most adequate system and provide such direct experimental evidence
in order to test the hypothesis of Ref. 2, which is important for
the reliability of the information extracted from PES experiment.

To study more thoroughly how the  sample resistivity affects the
line shapes of photoemission spectra, we chose cubic perovskite
\lsmaox ($\delta$ = 0, 0.03, and 0.06) systems. The transport
properties of Al-doped \lsmo were well studied by Y. Sawaki {\it
et al.}\cite{sawaki}. According to their reports, when a small
amount of aluminum is doped in \lsmo it substitutes B site (Mn)
element without any structural transition. Aluminum ion has the
closed-shell configuration with no $d$ electrons, so it gives the
same valence as the host manganese (+3) and no magnetic moments.
Hence aluminum doping induces only two kinds of changes. One is
the increase of random electrical potential, and the other is the
local cutoff of magnetic interaction between the spins of $t_{2g}$
electrons, {\it i.e.} it increases magnetic randomness. These
changes are enough to cause a resistivity change of two orders of
magnitude from $\delta$ = 0 to $\delta$ = 0.06. The useful
property most relevant to our study is the first one. Because the
dopant aluminum gives three electrons it does not change the hole
concentration of \lsmo. Thus the increase of resistivity with
increasing dopants is solely due to the increase of the carrier
scattering rates as a result of the increased randomness in
potential. This simplifies the analysis of the data in the model
of Ref. 2 dramatically. In most cases, there are too many
parameters to consider when a systematic study is undertaken to
understand how the sample resistivities affects the line shapes of
the photoemission spectra. For example, the temperature dependent
resistivity variation can come about not only from the change of
the scattering rate but also from the changes of the carrier
concentration or the electronic structure originating from phase
transition\cite{schulte}. (Recall in Drude model the DC
conductivity is given by $\sigma_0 = {ne^2 \tau}/m^* $, where $n$
is carrier concentration, and $\tau$ relaxation time, m$^*$
effective mass.) But in our case, all the parameters are fixed
except for the scattering rate of carriers (or inversely
relaxation time $\tau$).

According to the calculations in Ref. 2, photocurrent intensity $I(\omega, T)$ is given by
\begin{eqnarray}
I(\omega, T) & = & P_0(T)N(\omega)f(\omega) \nonumber\\
 &  & + \int^{\infty} _{0} P(\omega'-\omega, T) N(\omega')f(\omega') d\omega'
\end{eqnarray}
, where\cite{comment}
\begin{equation}
P(\omega) = \frac{4 \pi e^2}{\hbar v \omega^2} \frac{\mathrm{Re} \{ \sigma (\omega) \} }{|1+\epsilon(\omega)|^2}
\end{equation}
with the normalization condition
\begin{equation}
1 = P_0 + \int ^{\infty} _{0} P(\omega) d\omega.
\end{equation}
Here, $P_0$ is the zero energy loss probability, $P(\omega)$ the
probability of losing the kinetic energy $\hbar\omega$,
$N(\omega)$ the temperature independent density of states(DOS),
$f(\omega)$ the Fermi-Dirac distribution function,
$\sigma(\omega)$ the conductivity, $\epsilon(\omega)$ the
dielectric function. To apply these formulae to the manganites,
the  Drude model was used with additional parameter $r$ following
the approach of Ref.2 :
\begin{equation}
\epsilon(\omega) = \frac{4\pi i}{\omega} \sigma(\omega) = \frac{4\pi i}{\omega} (r+\frac{1}{1-i\omega\tau}) \sigma_0
\end{equation}
for  the dielectric function. Here $r$ represents the relative
strength of the frequency-independent part compared with Drude
part in the conductivity.  This parameter $r$ is somewhat
artificial and its physical origin is not yet well understood, but
inclusion of this parameter makes a sloping line shape often
observed in PES. In this paper we will assume that the above
scheme is suitable for describing the line shapes of PES for
manganites and test if it describes the experimental line shapes
of samples with different resistivities consistently.

\section{Experiment}
\label{sec:2} High quality polycrystalline specimens of
La$_{0.7}$Sr$_{0.3}$Mn$_{1-\delta } $Al$_\delta $O$_3$ ($\delta $
= 0, 0.03, 0.06) were made by the standard solid state reaction
method. Stoichiometric amounts of high purity ($\geq $ 99.99 \%)
La$_{\text{2}}$O$_{\text{3}}$, SrCO$_{\text{3}}$,
Al$_{\text{2}}$O$ _{\text{3}}$, and MnO$_{\text{2}}$ powders were
weighed and mixed with a pestle and a mortar. After calcining and
grinding repeatedly, resulting powders were pressed into a pellet.
Three pellets of La$_{0.7}$Sr$_{0.3}$Mn$ _{1-\delta }$Al$_\delta
$O$_3$ ($\delta $ = 0, 0.03, 0.06) were prepared together. And,
they were finally sintered at 1440$^{\circ }$C for 24 hrs and
slowly cooled in air. We applied the identical synthesis condition
for all the samples investigated. We cut each sintered pellet into
two pieces, one for transport measurement and the other for
photoemission spectroscopy. The resistivities were measured by the
conventional four probe method. The photoemission spectra were
taken with VG Microtech CLAM-4 multi-channeltron electron energy
analyzer with the energy resolution of 40~meV full width at half
maximum (FWHM) at Seoul National University under the base
pressure of $1.0 \times 10^{-10}$ torr. Photon source was
unmonochromatized He~{\footnotesize I} line ($h\nu$ = 21.2 eV).
The samples were cooled down to 95~K with liquid nitrogen and were
fractured {\em in situ} to obtain a clean surface at that
temperature by means of a top post. We obtained the spectra within
one hour after the cleave, and checked the surface contamination
by taking the valence band spectra.

\section{Results and Discussion}
\label{sec:3}

Figure~1 shows the angle integrated photoemission spectra up to 9
eV binding energy of \lsmaox for three doping cases $\delta$ = 0,
0.03, 0.06. Several prominent features can be seen in the spectra
and are labeled as A, B, C, D, and E as shown in the figure. Two
peaks A and B are strongly hybridized Mn $t_{2g}$ and $e_{g}$
states with O 2$p$ states. Peak C is usually assigned to O $2p$
nonbonding states and peak E to Mn $3d$- O 2$p$ bonding
states\cite{saitoh}\cite{cwlee}. Peak D around 4.5 eV in
$\delta$=0.03 and 0.06 cases is thought to originate from Al 3$sp$
impurities. The overall shapes are very similar to one another and
to those of previously published spectra for
$\rm{La_{0.7}Sr_{0.3}MnO_{3}}$\cite{fujimori} and
$\rm{La_{0.67}Ca_{0.33}MnO_{3}}$\cite{jhpark1}. But small doping
effects are clearly seen in several points. The position of peak E
is slightly shifted to higher binding energy side and a small peak
D appears around 4.5 eV as Al is doped. The apparent position
change of peak E is thought to come from the change of relative
concentration of $\rm Mn^{3+}$ and $\rm Mn^{4+}$ ions. Since $\rm
Al^{3+}$ ion replaces $\rm Mn^{3+}$ site, the spectral weight due
to $\rm Mn^{3+}$ sites will be reduced upon Al doping, which
should lie at lower binding energy than $\rm Mn^{4+}$ because of
correlation energy. The O $2p$ nonbonding states (peak C) show
almost no change, as expected. Mn $t_{2g}$ (peak B) and $e_{g}$
states (peak A), which are more intimately involved in the
transport properties of manganites, are also clearly seen around 2
eV and Fermi level and seem to be identical to undoped
case\cite{jhpark2}.  From these spectra we infer that the overall
electronic structures do not change much with doping and
consistent with the results of structural and transport study by
Y. Sawaki {\it et al.}\cite{sawaki}.  This is a strong evidence
that the increase of resistivity is not due to the change of
electronic structure or phase transition, but due to the increase
of scattering rates of carriers.  We note that for the doped
samples, specially $\delta$ = 0.06, the resistivity is fairly high
(see the inset of Fig~3) and the estimated electron mean free path
is comparable to the unit cell spacing\cite{comment2}. In spite of
this fact, resistivity curve and PES data show doped samples are
still metallic.  This behavior is similar to the cases of
A$_{3}$C$_{60}$ (A = K, Rb), La$_{1.85}$Sr$_{0.15}$CuO$_{4}$, and
Sr$_{2}$RuO$_{4}$\cite{bad metal}, where its physical origin is
currently under active debate.

To study the relation  between the line shape of PES near the
Fermi energy (${\rm E}_{\rm F}$) and the sample resistivity, we
obtained the photoemission spectra near the Fermi level in detail.
These spectra are shown in Fig.~2. Each spectrum is normalized to
the height at 0.6 eV below the Fermi level.  We can see clearly
the slight decrease of the spectral weight of "Mn $e_{g}$ band" as
Al is doped, which is as expected. Because the atomic cross
section of O $2p$ is about one order of magnitude larger than that
of Mn $3d$ at $h\nu$=21.2 eV\cite{lindau}, most of the spectral
weight change near ${\rm E}_{\rm F}$ reflects the O 2$p$
characters mixed in the Mn $e_{g}$ bands.  Even though the sample
resistivity changes by about two orders of magnitude, the whole
line shapes remain very similar to one another, and only small
change of PES spectra is detected.  This result seems to be quite
different from the expectation of the argument of Ref. 2, as
discussed below in more detail.

First of all, we must check whether  the resistivity of our
samples covers the valid region for the test of this theory.
According to a recent report\cite{joynt2}, there are two
inequalities that should be satisfied by samples for the extrinsic
effect to be important in PES.  For our experiments with the
values of the analyzer resolution R = 40 meV, band structure width
of interest B = 1 eV, carrier concentration\cite{asamitsu} $n =
1.5 \times 10^{22} cm^{-3}$ under the assumption that the ratio of
effective mass to bare mass of electron is order of 1, resistivity
$\rho$ must lie in the region from 1 to 100 $m\Omega cm$ for
"insulator regime" and $\sigma_{0}/\tau$ from $5 \times 10^{26}$
to $5 \times 10^{29} sec^{-2}$ for "metallic regime" to meet this
criterion.  Clearly, our samples pass through these regions (see
table I). So the fact that we observe nearly no change of PES line
shape in our spectra is quite surprising if significant extrinsic
effect does exist.

To compare the  experimental results and the theoretical
expectation based on the argument of Ref. 2, we try to fit the
spectrum of undoped sample with the theoretical formula (1) by
adjusting parameters $P_{0}$ and $r$ under the assumption of a
constant DOS as in Ref. 2. This comparison is thought to be a good
estimation on how reasonable the model calculation is. From this
fit, we determine the values of parameter $P_{0}$ and $r$ as well
as $\sigma_{0}/\tau$. Since $\sigma_{0}/\tau = ne^{2}/m^{*}$ in
the Drude model, we expect its value to remain the same for
Al-doped samples as well. In addition, there are no physical
reasons for $P_{0}$ and $r$ values to change significantly with
small Al-doping\cite{comment3}. Hence we calculate the line shape
of PES for the other two doped samples using formula (1) with the
same values for $P_{0}$, $r$ and $\sigma_{0}/\tau$, and compared
with the experimental spectra. The results are displayed in Fig.~3
with the label of fitting 1. Here the calculated lines are
convoluted with the experimental resolution of 40 meV in FWHM and
the height of spectra are adjusted to coincide with the experiment
around 0.6 eV below the Fermi level.

For $\delta$ = 0 case,  we are able to fit the experimental
spectrum quite well with the line shape calculated from eq. (1) by
proper choice of $P_{0}$ and $r$ values. The resulting fitting
values are similar to those in Ref. 2 except the zero energy loss
probability $P_{0}$. Though $P_{0}$ in our fitting ($P_{0}$ =
0.05) is much larger than that of Ref. 2 ($P_{0} \leq $ 0.0025),
it is still too small in comparison with the estimated value from
the sum rule derived by D. L. Mills\cite{mills} ($P_{0}$ = 0.35)
or electron energy loss spectra (EELS) taken by K. Schulte {\em et
al.}\cite{schulte} ($P_{0}$ = 0.82). For $\delta$ = 0.03 and 0.06
cases, we can see that the photoemission line shapes calculated
from eq. (1) with the same parameter values show severe
discrepancies from those of experimental spectra.  Note the
behavior of resistivity dependence of calculated PES line shape.
Even the small change of resistivity around 1 $m\Omega cm$ is
expected to make large effect on the line shape. Other parameters
give relatively weak dependence.  This strong resistivity
dependence of PES line shape in eq (1) suggests the possibility
that the PES can give severely distorted information about the
density of states near Fermi level when the material under
investigation is a poor conductor like manganites, if the
extrinsic ohmic loss effect is indeed important.

The zero energy loss probability $P_{0}$  is assumed the same for
all samples in fitting 1. In fitting 2, $P_{0}$ is also allowed to
vary depending on $\delta$-value to see if this gives better
agreement with experiment, although we do not think $P_{0}$ should
be so much dependent on sample resistivity. However, we see that
the results of fitting 2 do not give significantly better
agreement between theory and experiment than the case of fitting 1
even though we take $P_{0}$ as free parameter for each $\delta$.
All the parameter values used in the above analyses are tabulated
in Table~I.

Next, we try to fit our data with the sloping DOS near ${\rm
E_{F}}$. In this fit, we do not use the additional parameter $r$
because using both sloping DOS and $r$ makes the analysis too
arbitrary. Figure~4 shows the results in detail. In this case, we
can fit all the experimental spectra quite well, but the obtained
parameter values seem to give another question. There is a big
difference between the values of $P_{0}$ in doped sample and
undoped ones. If we use a small value of $P_{0}$ for undoped
sample, the calculated spectrum always shows the step-like feature
around 400 meV binding energy. This is the effect of surface
plasmon\cite{joynt}. To remove this step-like feature in the
calculated spectrum, $P_{0}$ value must be larger than 0.5 at
least as seen in Fig~4. This abrupt change of $P_{0}$ value by
more than one order of magnitude with small Al-doping is not
expected judging from the EELS study of alloy semiconductor ${\rm
Al}_{x}{\rm Ga}_{1-x}{\rm As}$\cite{thiry}, and cannot be
explained even if we include the multiple scattering process.

From our experimental results and fittings as stated above, we
conclude that the experimental spectra do not show any symptom of
significant ohmic loss effect, and the theoretical line shapes
cannot reproduce experimental spectra for our series of manganite
samples in a reasonable way without much conflict.  Hence it seems
clear that the long range Coulomb interaction between the outgoing
photoelectrons and the sample left behind is not so large as was
suggested in Ref. 2. Our fitting results show that the only way to
explain the experimental spectra consistently is to increase the
$P_{0}$ value rather large, and this means that we can regard the
PES spectra near Fermi energy as the replica of the density of
states in condensed matter whether its conductivity is good or
not.

\section{Conclusion}
\label{sec:4}

In summary,  we presented an experimental evidence that the theory
suggested in Ref. 2 cannot explain consistently the behaviors of
photoemission spectra for a series of systems in which
resistivities vary systematically. Hence we conclude that the
influence of this long range Coulomb interaction is not so large
as can distort the photoemission line shape significantly.

\acknowledgements
This work was supported by the Korean Science and Engineering Foundation through
Center for Strongly Correlated Materials Research (CSCMR) at Seoul National University.
KHK was also supported by the BK-21 Project of the Ministry of Education.

\begin{figure}
\begin{center}
\epsfig{file=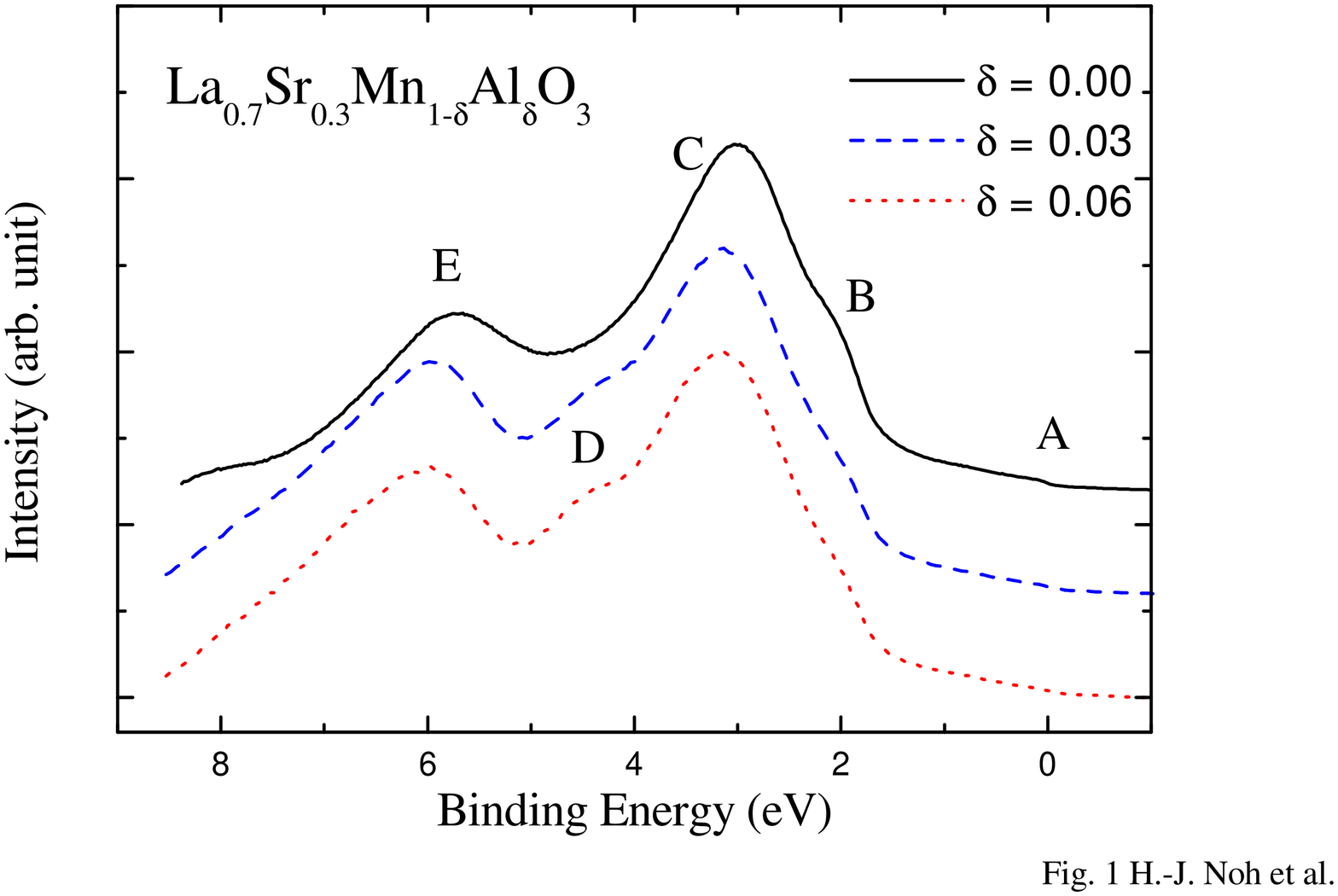,width=9.0cm,angle=0.0}
\end{center}
\caption {Valence band photoemission spectra of \lsmaox.
Solid, dashed, dotted lines are for $\delta$ = 0, 0.03, 0.06 samples respectively.
All the spectra are taken using He~{\footnotesize I} line (h$\nu$ = 21.22 eV) at the temperature 95~K.
}
\end{figure}

\begin{figure}
\begin{center}
\epsfig{file=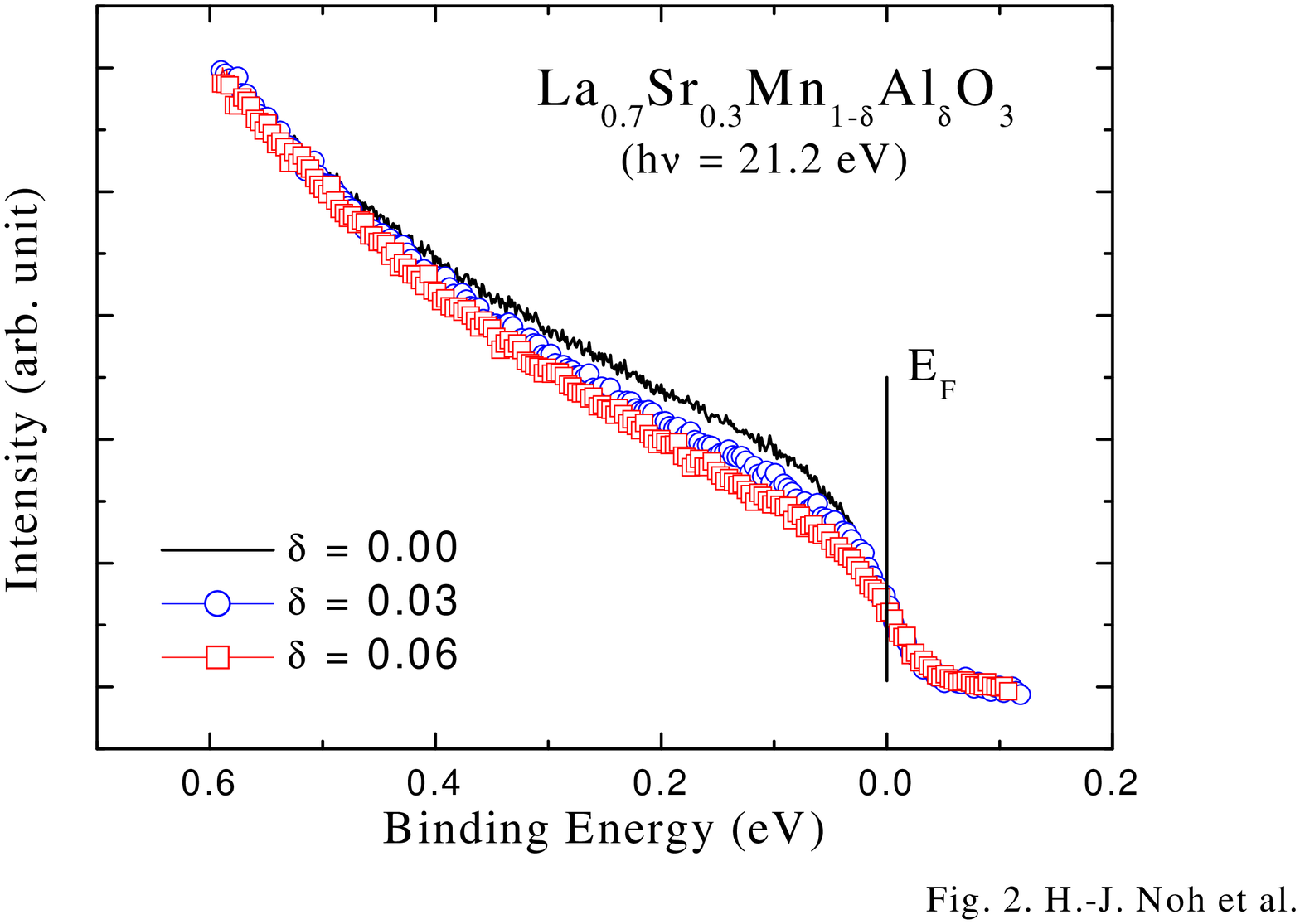,width=9.0cm,angle=0.0}
\end{center}
\caption {Near ${\rm E}_{\rm F}$ photoemission spectra of \lsmaox.
Solid line is for $\delta$ = 0, solid circles for $\delta$ = 0.03, solid squares for $\delta$ = 0.06.
Vertical line is the Fermi level determined by the reference Au sample.
}
\end{figure}

\begin{figure}
\begin{center}
\epsfig{file=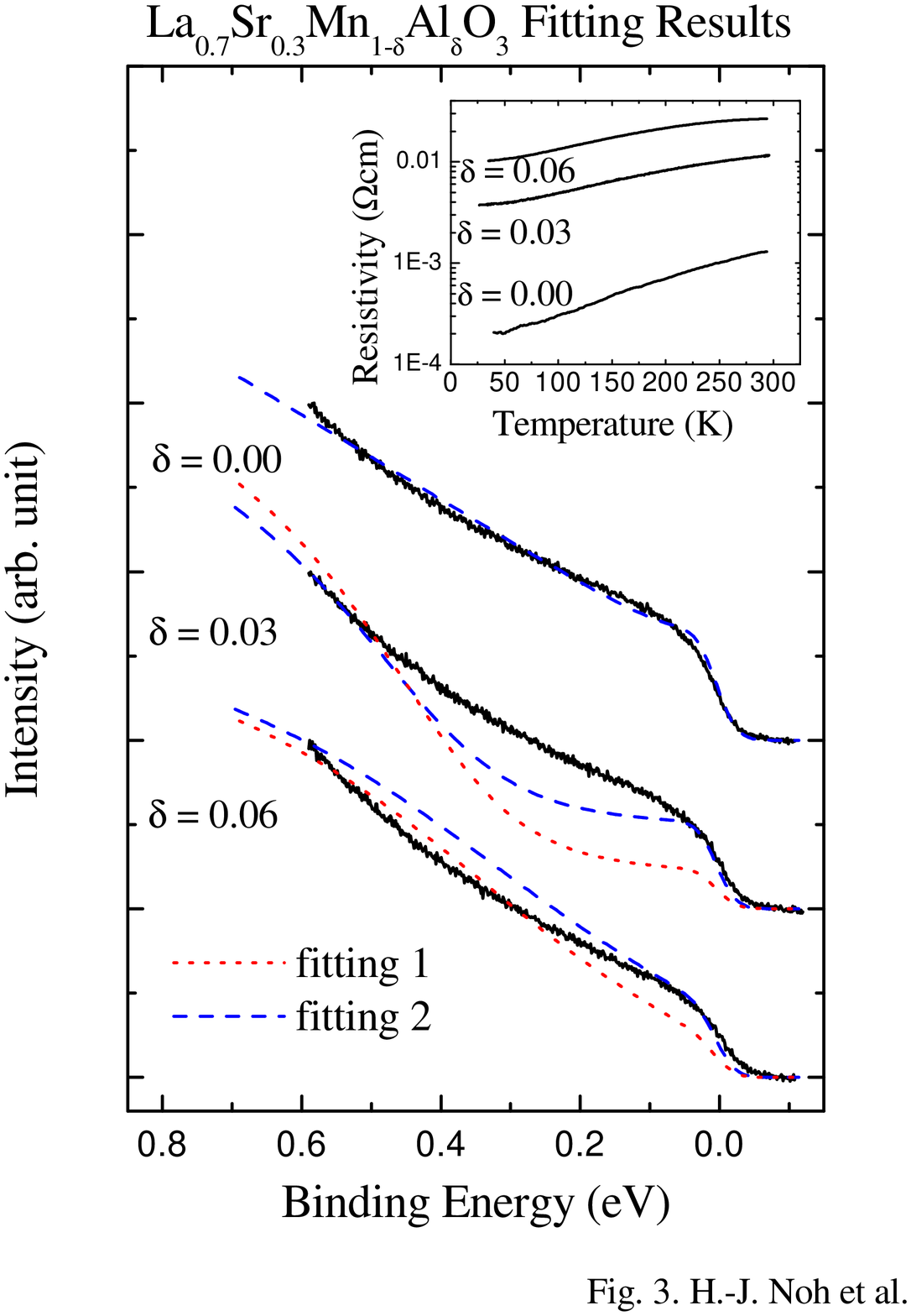,width=9.0cm,angle=0.0}
\end{center}
\caption {Experimental vs.  calculated fitting spectra. The
parameter values used in the fitting are listed in Table~I. In
fittings 1 and 2, all the parameter values are the same except for
the zero energy loss probability $P_{0}$ (see Table~I). Inset:
Resistivity vs. temperature curve of \lsmaox samples used in our
PES experiments. }
\end{figure}

\begin{figure}
\begin{center}
\epsfig{file=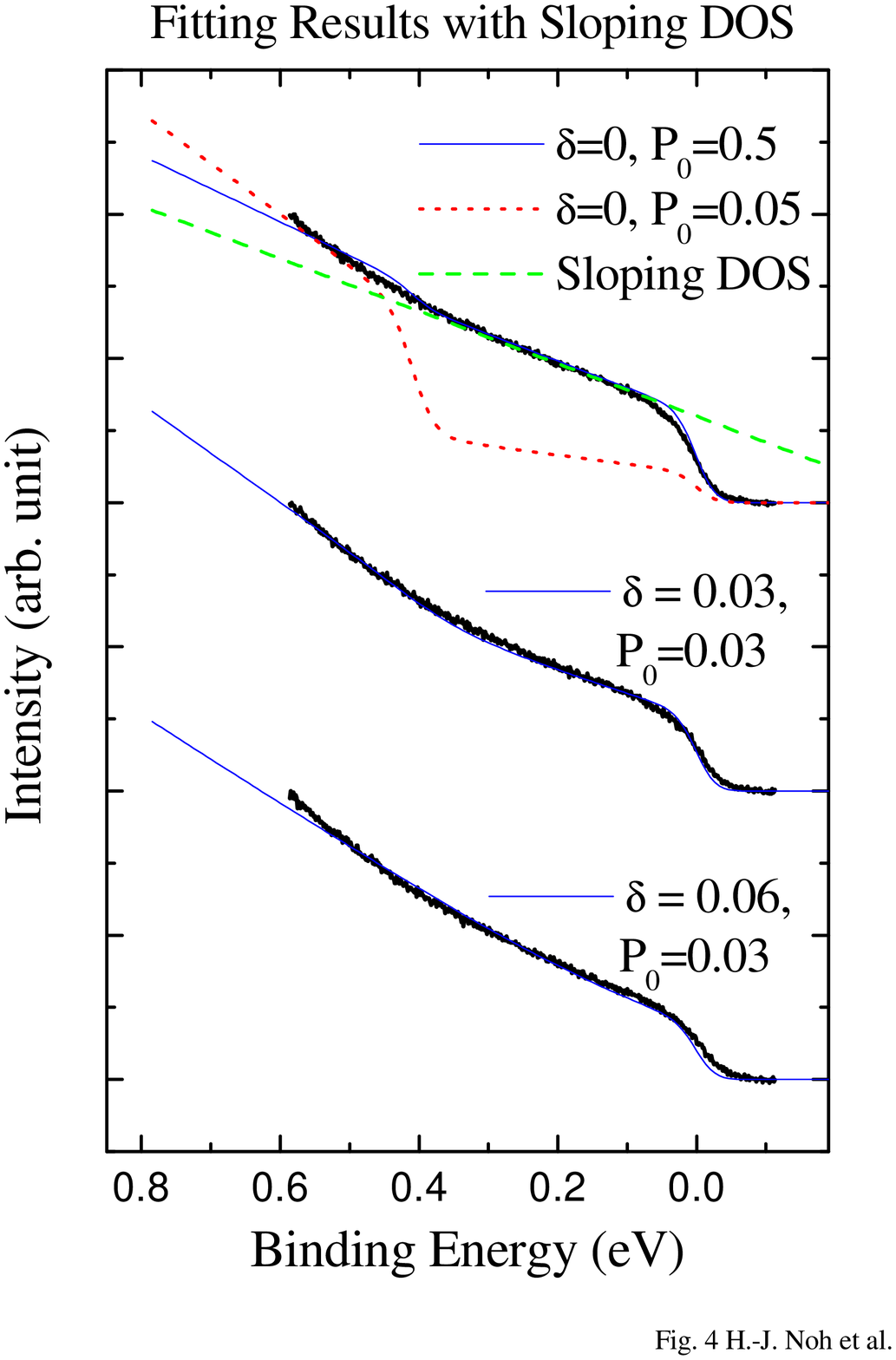,width=9.0cm,angle=0.0}
\end{center}
\caption {Experimental vs.  calculated fitting spectra with
sloping DOS. For each experimental spectrum, the same sloping DOS
is used as shown with the dashed line in $\delta$ = 0 case. }
\end{figure}

\begin{table}
\caption { Parameter values in the analyses. The resistivities of
each sample are taken at 95~K.  ${\sigma_{0}}/{\tau}$  should be
the same for all samples in the Drude model, since it is equal to
$ne^{2}/m^{*}$. $r$ and $P_{0}$ values in fitting 1 are fixed for
all samples as those obtained from the fitting of undoped sample
spectrum, while in fitting 2 $P_{0}$ is allowed to vary depending
on the sample. }
\begin{tabular}{c | c c c c c c}
$\delta$ & $\rho$ & ${\sigma_{0}}/{\tau}$ & $r$ & $P_{0}$ & $P_{0}$ & $P_{0}$ \\
         & (m$\Omega$cm) & (1/sec$^2$ )   &     &(fit 1)&(fit 2) & (sloping DOS)\\
\hline
0.00 & 0.29 & $6.08 \times 10^{28}$ & 0.25 & 0.05 & 0.05 & 0.5\\
0.03 & 4.8 & " & " & "  & 0.12 & 0.03 \\
0.06 & 13.0 & " & " & " & 0.10 & 0.03 \\
\end{tabular}
\end{table}

\end{multicols}

\begin{references}
\bibitem {comment} The correct prefactor was derived by D. L. Mills\cite{mills}. In this paper, because $P_{0}$ is
a free parameter, calculated line shape does not depend on the absolute value of the prefactor.
\bibitem {comment2} The resistivity of polycrystalline manganites is always higher than that of single crystials
due to the intergrain tunneling effect.\cite{KHK}. To estimate the
mean free path, intragrain conductivity must be used, which can be
obtained from single crystal sample\cite{sawaki}. Under the
assumption of a spherical Fermi surface, the mean free path $l$ is
given by  $l = \hbar k_{F} \sigma_{intra} / ne^{2}$ if the Drude
formula is used. Taking $n = 1.5 \times 10^{22} cm^{-3}$
\cite{asamitsu}, we obtain $l = 2.25/\rho_{m\Omega cm}$ \AA. Here,
$\rho_{m\Omega cm}$ means the resistivity in $m\Omega cm$ unit. If
we take $\rho_{m\Omega cm}$ = 0.3 for $\delta$ = 0.06 sample, $l$
= 7.5 \AA, which is close to the unit cell spacing $a$ $\sim$ 8
\AA.
\bibitem {comment3} If we consider multiple scattering process, the sum rule\cite{mills} does not hold, and
$P_{0}$ can change with Al doping because the dopant Al cause the
change of dielectric constant in our system through the change of
relaxation time. However, its dependency cannot be strong under
the condition of no structural transition and constant carrier
concentration, which is consistent with EELS experiments on ${\rm
Al}_{x}{\rm Ga}_{1-x}{\rm As}$ \cite{thiry}. In the case of $r$
value, it represents the incoherent part of optical conductivity,
which is almost constant below 0.1 eV in infrared spectroscopy
data. There is no clear reason why this incoherent part must be
affected by the increase of scattering rates of carriers.
Furthermore, many results of optical conductivity of
manganites\cite{jhjung} indicate that the incoherent part will
remain almost the same with small Al doping since the electronic
structure does not change and the carrier concentration and
temperature are fixed in this study.
\end{references}
\end{document}